# Micromagnetic Design of Bias-Free Reconfigurable Microwave Properties in Hexagonal Shaped Multilayer Nanomagnets

Krishna Begari

*Department of Physics, Indian Institute of Technology Tirupati, Tirupati, India*

Email ID: *krishnabegari.phy@gmail.com*

**Abstract**

Magnetic miniaturized nanostructures hold great promise for current and future microwave technologies due to their magnetization dynamics in the GHz frequency range. This work presents a method for investigating reconfigurable microwave properties using a novel hexagonal nanomagnet structure. Micromagnetic simulations are employed to investigate the magnetic static and dynamic properties of the nanomagnets. A simple field initialization method is used to examine two distinct magnetic remanent states in each sample. A nanosecond-width magnetic pulse field can be applied to tune the unique magnetization dynamics parameters corresponding to the different remanent states. Find that for both single-layer and multilayer nanomagnets, there is a notable frequency shift in the sub-GHz and GHz regions between the two distinct magnetic remanent states.

I. INTRODUCTION

Magnetic materials are the vital elements of standard microwave magnetic devices, and they are very substantial in size [1]. Therefore, device miniaturization is increasingly important in current and emerging technologies. Recently, several studies have reported that miniaturized nanostructures and their arrays operate in the GHz (microwave) frequency range, showing great potential for future microwave device miniaturization and integration [2–12]. A significant benefit of microwave magnetic devices over their electronic or photonic counterparts is their convenient tunability utilizing external variables, such as laser current induced Oersted field [13], heating [14], magnetic field [15–17] and Heated point of a scanning microscope[18]. For example, by adjusting the configuration and bias magnetic field ($H_{ex}$) nanomagnetic arrays referred to as magnonic crystals—can exhibit tunable ferromagnetic resonance spectra [5]. It is important to note that device integration at the microscale is not suitable for tunability approaches that rely on an external bias magnetic field. Hence, nanoscale magnetic structures capable of tunability absence of magnetic field are crucial for the practical realization and on-chip magnetic nanodevices [19]. Although, only a few studies have demonstrated without any external bias magnetic field based on nanostructures like arrow-shaped nanomagnets [20], rhomboid-shaped nanomagnets [21], nanowires [7, 22, 23], dipolar-coupled nanopillars [24, 25], and rhomboid shaped multilayer nanomagnets [26]. These devices operate based on multiple remanent states, each corresponding with different microwave properties. Intelligent nanomagnetic structures and their arrays or networks are becoming increasingly significant, as the upcoming generation of integrated microwave devices requires ultra-fast operation with ultra-low power consumption.

This report presents the magnetic and microwave properties of a newly designed hexagon-shaped nanomagnet. Reconfigurable microwave behavior is demonstrated without any external bias field ($H_{ext}$) in three types of magnetic nanostructures: (A) single layer nanostructures and their dipolar-coupled networks, (B) an isolated multilayer nanomagnet, and (C) dipolar-coupled multilayer nanomagnets. Micromagnetic simulation techniques are employed to investigate magnetization reversal processes and microwave characteristics in detail. This approach serves as a powerful tool for evaluating proof-of-concept designs before committing to costly nanofabrication and yields results consistent with experimental observations. I observed two distinct remanent magnetic states, each exhibiting unique microwave properties within the 3–16 GHz frequency range. These findings highlight the potential for low power consumption and rapid tunability, which are critical for the advancement of next generation nanoscale microwave devices.

II. METHODS

All results were obtained using micromagnetic simulations performed with the Object Oriented Micromagnetic Framework (OOMMF) software [27]. It is a finite difference method (FDM) for solving differential equations followed by approximating derivatives with finite differences. We have a differential equation is Landau-Lifshitz-Gilbert (LLG) differential equation which is contribution of contains precision and damping terms of magnetization in the equation. In FDM both spatial domain and time intervals are broken into several steps. To evaluate magnetization dependent on both temporal and spatial variables, the LLG equation [28] is used.

$$\frac{dM}{dt} = -\gamma M \times H_{eff} - \lambda M \times (M \times H_{eff})$$

$$\lambda = \alpha \frac{\gamma}{M_S}$$

where $\gamma$ is the gyromagnetic ration of electron and α indicates dimension less damping parameter. Here, magnetic effective field ($H_{eff}$) is contributed from the applied magnetic field, demagnetizing field and some other effect fields. The first term belongs to precision and the second term from the damping of magnetization. The magnetic and microwave properties are demonstrated following static and dynamics simulations, respectively.

Simulation parameters of the permalloy sample ($Ni_{80}Fe_{20}$ alloy) are magnetization ($M_S$) = 800 emu/cm$^3$, exchange stiffness (A) = 13×10$^{-7}$ erg/cm, and magneto crystalline anisotropy (K) = 0. The sample was discretized into no. of cubic cells in the FDM. The exchange length sets an upper limit for the cell size (5.7 nm) of permalloy for more accurate results and have considered it as 5 nm along x, y and z axis for every cubic cell. The magnetic hysteresis loops, magnetic remanent states and dipolar field distribution have been demonstrated using static micromagnetic simulation considering high damping constant alpha value (0.5) to optimize the simulation time. The field step 10 Oe considers generating the magnetization reversal. The dynamics simulation has been accomplished using normal damping constant (0.008). To get the FMR spectra, a sinc pulse field $(H_S) = H_0 sinc(2\pi f_c \tau)$ has been applied along x-axis to the remanent magnetic states. where $H_0 = 50\ Oe$, $f_N$ is Nyquist frequency greater than cut-off frequency($f_c$) and $\tau$ refers to the simulation time. The time dependent magnetization recorded at every 10 picoseconds up to 4 nano seconds period. Ferromagnetic resonance spectra obtained by performing FFT on the time dependent magnetization data ($m_x$). 2D profiles of Ferromagnetic resonance modes carried out from an analysis of magnetization in time and space *M (t, r)*.

### III. RESULTS AND DISCUSSION

    A. Single layer Hexa-Shaped Nanomagnets (SL-HNM)
    B. Isolated Multilayer Hexa-Shaped Nanomagnet (ML-HNM)
    C. Multilayer Network Systems (ML-HNM-N)

D. Ultrafast switching

## A. (i) Investigation of remanent states in single layer HNM and networks

Nanostructures have been designed based on hexagonal geometry structure with dimensions are width (w)= 130 nm, thickness (t)= 25 nm , and length (L)= 450 nm. Permalloy serves as a ferromagnetic layer for all the samples in this case. Single layer isolated nanomagnet, single layer dipolar coupled two nanomagnets, single layer dipolar coupled three nanomagnets, single layer dipolar coupled four nanomagnets labelled as SL-1M, SL-2M, SL-3M and SL-4M, respectively.

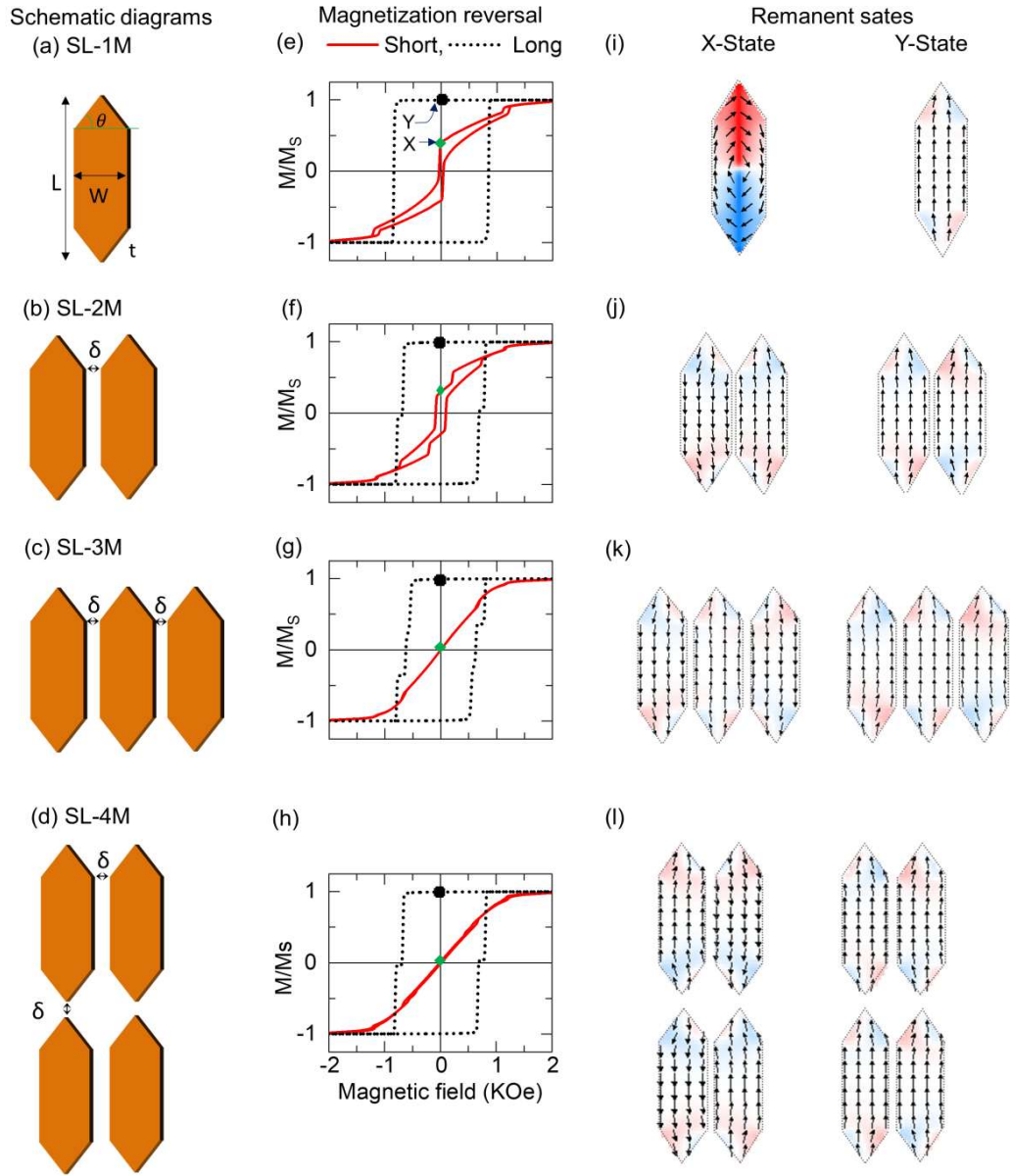

**Figure 1. Single-layer hexagonal nanomagnets (HNMs) and their dipolar-coupled configurations.** (a–d) Schematic representations of the nanomagnet geometries, (e–h) simulated magnetization reversal processes under an applied external magnetic field and (i–l) depicts the corresponding remanent magnetic states and associated spin textures.

The interlayer spacing ($\delta$) of the dipolar coupled nanomagnets is 30 nm. Which is less than the width of the nanomagnets therefore their having dipolar coupling between them. Figure 1(a-d) shows schematic diagrams of the single layer nanomagnets with separation ($\delta$). The magnetization reversal obtained using micromagnetic simulations by applying an external magnetic field along the x-axis and y-axis for SL-1M, SL-2M, SL-3M and SL-4M as shown in figure 1(e-h). The saturation field found around ($\pm$) 800 Oe in the y-axis M-H loop for all the samples. Multi-step magnetization reversal is observed in the x-axis M-H loop for SL-2M while decreasing the field a long x-axis. These steps in M-H loop appear due to magnetic re-orientation at edges of the sample and anti-parallel configuration with the nearest nanomagnet. However, the single magnetization loop along x-axis was observed in SL-3M and SL-4M samples. Two remanent magnetic states investigated followed by simple field initialization process. When field initialized along the x-axis then removed followed by 2000 Oe to Oe and observed the magnetic state called as remanent magnetic state donated with X-state (at $H_{ext}$=0). The X-state ($\blacklozenge$) has vortex domains in the SL-1M. Field initialized along y and removed it to get another magnetic configuration state which is denoted as 'Y-state ($\blacksquare$)'. Figure 1(i) shows X-state has vortex domains and single domain existing for the Y-state in the SL-1M. Figure 1(j-l) shows remanent magnetic X and Y-states of single layer dipolar coupled network nanomagnets have anti-parallel and parallel magnetic configuration, respectively. These two different magnetic states have distinct dynamics properties. Thus, by utilizing various mechanisms, this study presents a demonstration of tunable magnetization dynamics responses.

## A. (ii) Zero-field reconfigurable microwaves in SL-HNM

The demonstration of magnetization dynamics of two different remanent states have been studied based on their Ferromagnetic resonance (FMR) spectra at absence of any biased field. Figure 2(a-c) shows the FMR spectra of X-state and Y-state for SL-1M, SL-2M, SL-3M, and SL-4M, respectively. The frequency responses observed in the FMR spectra range from 3 to 16 GHz, covering both the X and C bands of the microwave spectrum. In general, the X-band spans a frequency range of 8 and 12 GHz, while the C-band spans 4 to 8 GHz in the microwave spectrum. A comprehensive 2D spatial analysis was conducted to determine the frequency modes in the FMR spectra of the X and Y states, as highlighted in Figure 2(d–f). Two types of localized modes were observed in the nanomagnets which is marked with two different of symbols 'solid triangular' ($\blacktriangle$) and 'solid circular' ($\bullet$). In addition, solid square ($\blacksquare$) nodes appeared in the magnetic remanent states. The '$\blacktriangle$' mode localized at the edges of the nanomagnets and '$\bullet$' mode existing at center of the nanomagnets. Solid square mode is found to be a nodal line for few remanent states. However, '$\blacksquare$' modes didn't found in X state of SL-1M sample. Interestingly found the clearer frequency shift is observed between X and Y-states of SL-1M and SL-4M samples. For instance, in the SL-1M system, the frequency modes $\blacktriangle$ and $\bullet$ were observed at 4.83 GHz and 8.54 GHz for the X-state, while the same modes appeared at 6.03 GHz and 9.34 GHz for the Y-state. The '$\bullet$' mode is considered as the central mode of both X and Y states for all samples. The frequency shift is calculated based on center mode position between X and Y states in all. The frequency shift of center mode found to be 800 MHz, 300 MHz, 618 MHz and 200 MHz for SL-1M, SL-2M, SL-3M and SL-4M, respectively. In single-layer samples, the frequency shift remains within the MHz range, attributed to the weak stray filed strength between adjacent nanomagnets.

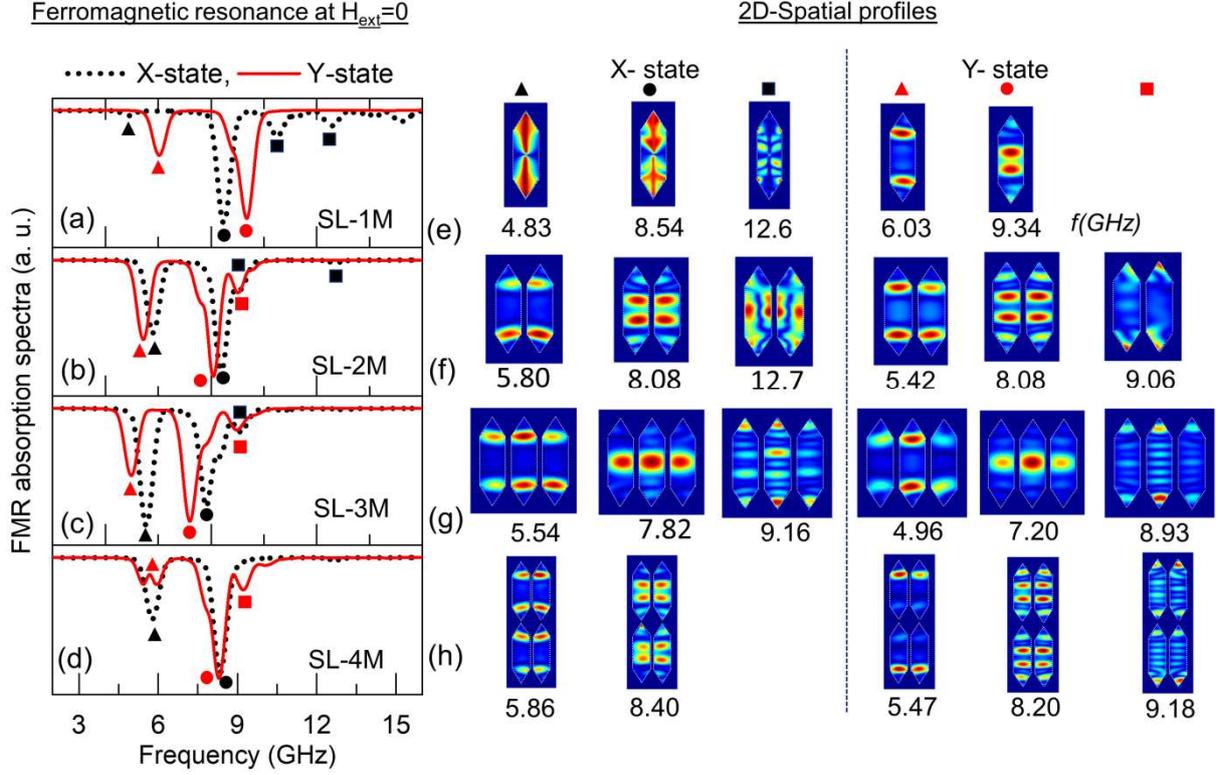

**Figure 2.** (a) Simulated ferromagnetic resonance (FMR) spectrum showing the frequency response of the system under an oscillating magnetic field and (b) two-dimensional spatial profiles corresponding to the resonant frequency modes.

## A. (iii) Origin of the Ferromagnetic resonance shifts in single layer samples

To understand the FMR shift between two distinct remanent magnetic states, the variation of the demagnetization field in single-layer samples was analyzed. The schematic of scan line of though the sample SL-1M as shown in Figure 3(a). Figure 3(b) displays the 2D profiles of the variation of demagnetization fields for the X and Y-states of sample SL-1M. Plotting line $L_1$ and $L_2$ scans along geometrical long axis and short axis of the sample as shown in figure 3(c & d), respectively. The dipolar field was quantitatively determined based on this analysis. The Kittel equation [29] used to estimate the FMR shift:

$$f_{res} = \frac{\gamma}{2\pi}\sqrt{\left(H_{eff} + (N_z - N_y)M_y\right)\left(H_{eff} + (N_x - N_y)M_y\right)},$$

where, γ is the gyromagnetic ratio ($\frac{\gamma}{2\pi}$ = 2.8 MHz/Oe), fres is the FMR frequency, Heff is the effective field and $N_x$, $N_y$ and $N_z$ denote the demagnetization coefficients, with their sum equaling 4π ($N_x + N_y + N_z = 4\pi$) in the CGS system.

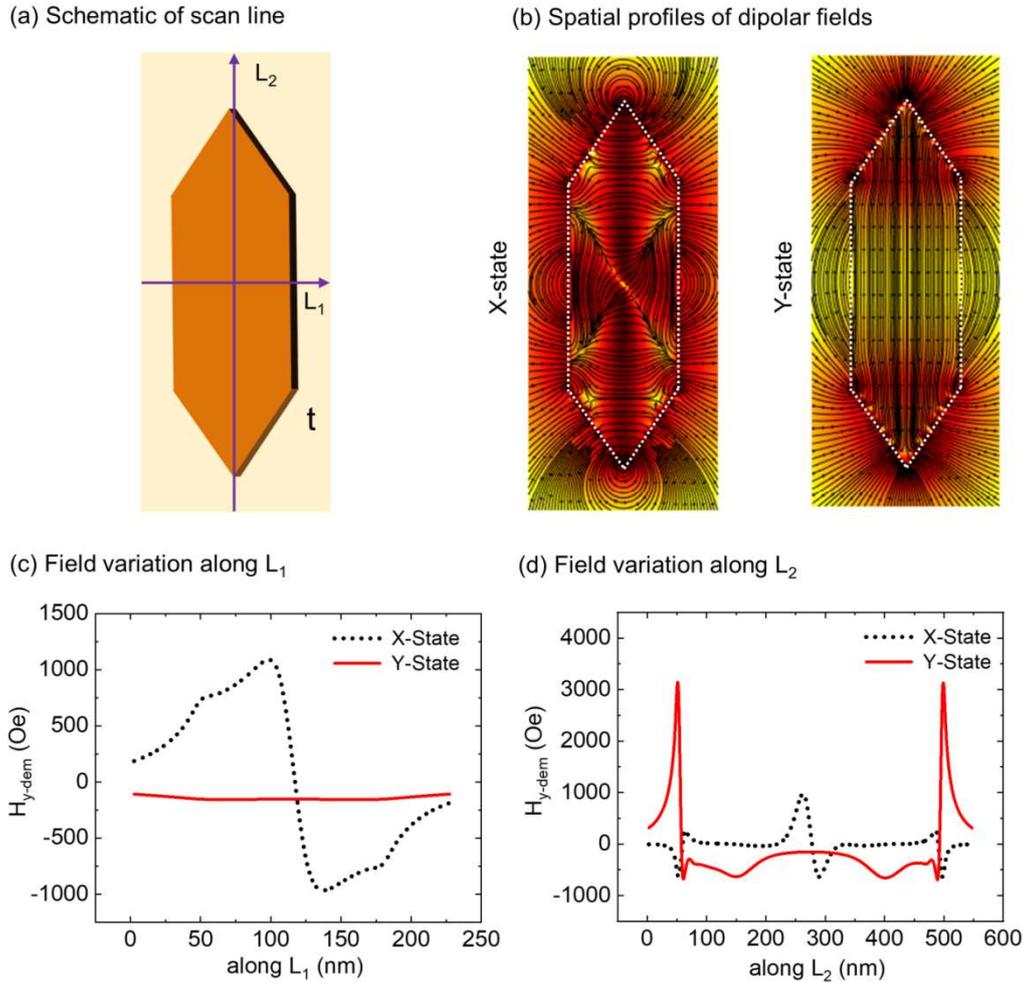

**Figure 3. Dipolar field distribution in a single-layer hexagonal nanomagnet.** (a) Schematic showing the scanning lines used for field analysis, (b) two-dimensional map of the demagnetization field and (c, d) line profiles of the dipolar field variation along the x- and y-axis of the nanomagnet, respectively.

$H_{eff} = H_{app} + H_d$ is the effective field that includes contributions from the stray field ($H_d$) and the applied field ($H_{app}$). To estimate the FMR shift in the present work, remanent states—where the applied field ($H_{app}$) is absent—were taken into account. Consequently, the two distinct remanent states' dipolar field changes should predict the effective field's ($H_{eff}$) variation. It is important to note that the Kittel equation applies to uniformly magnetized materials, which does not hold valid for the HNM samples at remanence. This analytical expression is applied solely to estimate the shift in FMR (Δf) due to specific dipolar field variations, rather than to determine the precise resonance frequencies of the modes in these specimens. To estimate the demagnetization factors $N_x$, $N_y$, and $N_z$, Aharoni's expressions [30] were applied under the assumption of a rectangular nanomagnet. The current sample geometry approximates a rectangular prism, making this idealization suitable for basic quantitative comparisons with analytical results. For dimensions of 450 nm in length, 130 nm in width, and 25 nm in thickness, the corresponding demagnetization factors are approximately 2.2316, 0.6135, and 9.7212 for $N_x$, $N_y$, and $N_z$, respectively. By using these numbers in the Kittel equation, calculated that the 800 MHz frequency shift observed in the instance of HNM-SL-1 requires a field fluctuation of approximately 151 Oe. As shown in Figure 3(c & d), the demagnetization field in the center region of HNM-SL-1M varies significantly from –33 Oe to –154 Oe ($\Delta H$ = 121 Oe) along L1 and L2 for the X and Y magnetic states. The analytical equations, though based on a simplified model, clearly show the main physical reasons behind the FMR shift and give results that agree with simulations. In the same way, the smaller differences in demagnetization field values between the two remanent states help explain the smaller FMR shifts in HNM-SL-2M and HNM-SL-3M compared to HNM-SL-4M (see Supplementary Figure S1).

**B. An isolated multilayer hexagon shaped nanomagnet (ML-1M)**

**B. (i) Two distinct remanent states in an isolated multilayer nanomagnet**

The magnetic properties of have been investigated based on multilayer hexagon nanomagnetic with dimensions width (w) = 130 nm and length (L) = 450 nm. As we know the strong dipolar field gives the large frequency shift between the two magnetic remanent stated in the multilayer nanostructures [22]. To address this, an isolated ML-HNM nanomagnetic structure was designed, consisting of two magnetic layers—top and bottom—separated by a thin, non-magnetic spacer layer. The spacer layer has a thickness of 10 nm, while the top and bottom magnetic layers are 15 nm and 25 nm thick, respectively. The schematic and dimensions of multilayer isolated nanomagnet are shown in Figure 4(a). Static magnetic properties carried out using damping constant alpha ($\alpha$) = 0.5 and results shown in figure 4(b). The solid hysteresis loop indicates when a field is applied along the short axis to the ML-1M and the dotted line belongs to the long axis M-H loop. Two distinct magnetic remanent states are indicated by the symbols (♦) and (■) on the M–H loops measured along the short and long axes, respectively. These magnetic remanent states X and Y can be obtained with field initialization process from 2000 to O Oe along x (X-state) and along y (Y-state) as shown in Figure 4(c).

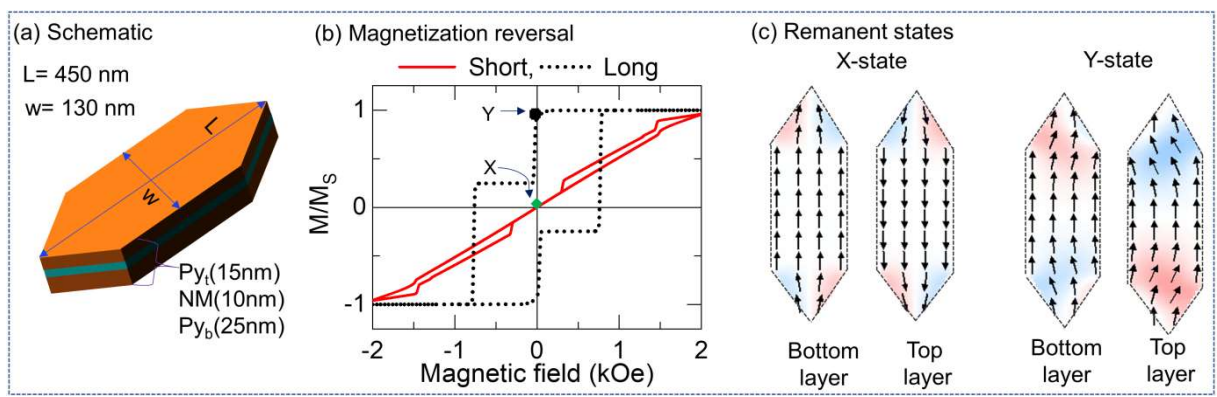

**Figure 3. An isolated multilayer nanomagnets.** (a) Schematic (b) hystereis and (c) remanent magnetic states in both the layers.

Distinct magnetic behaviors were observed for the X-state and Y-state, corresponding to anti-parallel and parallel magnetization configurations, respectively. In the X-state, the magnetization initially relaxes into a uniform direction; however, due to strong dipolar coupling, the magnetization in the thinner layer (top layer) reverses, resulting in an anti-parallel configuration. This behavior arises from the significant dipolar interaction between the top and bottom magnetic layers, which are separated by a thin 10 nm non-magnetic spacer. The anti-parallel configuration also aids in reducing the stray field through flux closure. In contrast, when the external field is applied along the Y-axis, the magnetization in both layers relaxes and remains aligned in the same direction, forming a parallel configuration. This occurs due to the dominant influence of shape anisotropy over dipolar interactions in this orientation. Thus, the Y-state exhibits a stable parallel magnetic configuration. The magnetization dynamics associated with these remanent states are discussed in the following section.

**B. (ii) Microwave properties in isolated multilayer nanomagnets at $H_{ext}$ = 0**

The microwave characteristics of an isolated multilayer nanomagnet were investigated through micromagnetic simulations by analyzing the FMR spectra. Figure 5(a) presents the FMR spectra corresponding to the X and Y states of an isolated multilayer hexagonal nanomagnet. The spectra were observed within the 3–16 GHz range without the application of any external bias field, covering both the X and C bands of the microwave spectrum. The observed clear prominent frequency modes with different frequency position as the results found the frequency shift between two distinct magnetic remanent X- and Y-states.

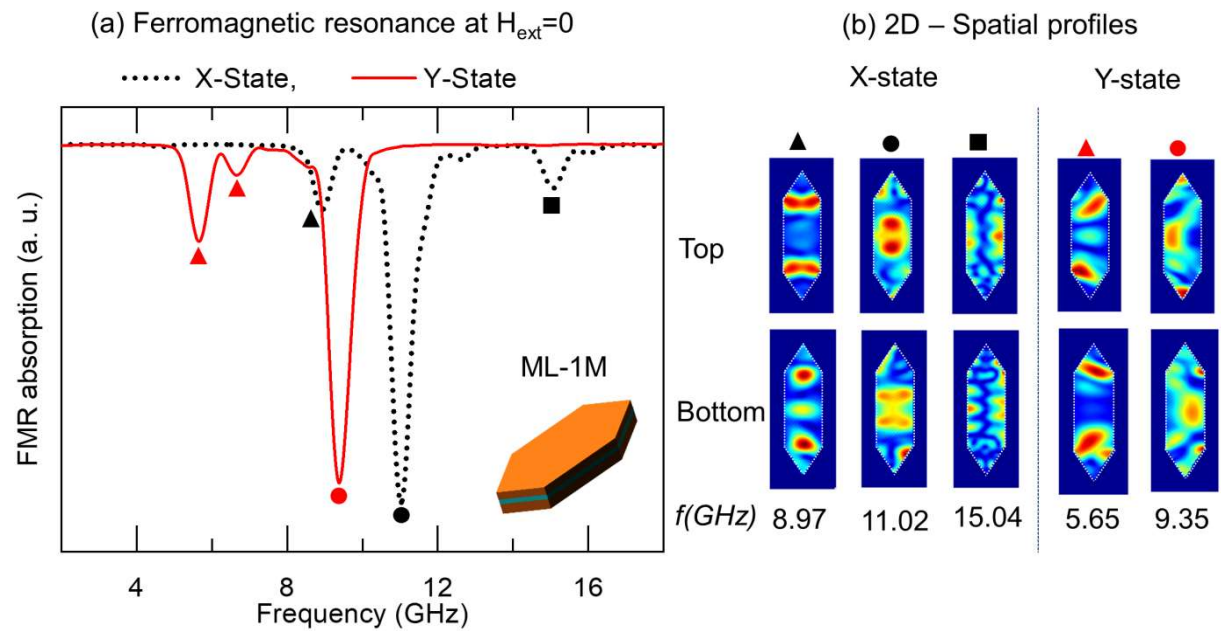

**Figure 4. Simulated ferromagnetic resonance (FMR) response of an isolated multilayer nanomagnet.** (a) Frequency-dependent resonance peaks showing the system's dynamic magnetic behavior, and (b) two-dimensional spatial profiles corresponding to the resonant modes, illustrating the distribution of magnetization dynamics within the structure.

To identify these prominent modes, I did a further 2D spatial profile analysis and shown in Figure 5(b). Here, two different types of frequency modes '★' and '●' observed in X and Y states. Additionally, a '■' mode exists in the X-state. The triangular mode and circular modes appear at the edges and center of the nanomagnets in the top and bottom layer. The most prominent frequency mode labeled with (Solid Square) found to be a nodal line of the multilayer ML-1M nanomagnet. These prominent modes (triangle, circle and square) have the frequency at 8.97 GHz, 11.02 GHz and 15.04 GHz for X-state, other hand; it will appear at 5.65 GHz and 9.35 GHz for Y-state. The central mode was considered for calculating the frequency shift between the two distinct remanent states. A significant shift of 1.7 GHz was observed, which is notably larger than the 0.8 GHz shift in the single-layer sample. This enhanced shift is attributed to the strong dipolar interaction between the two magnetic layers in the isolated multilayer nanomagnet.

**B. (iii) Estimation of dipolar fields in an isolated multilayer nanomagnet**

To comprehend the source of this significant FMR change, I calculated the variations in dipolar fields within the multilayer structures. For this investigation, I selected the ML-1M multilayer sample to evaluate the dipolar fields in both the X and Y states. Figure 6(a&b) shows the demagnetization field distribution and 2D spatial profiles for the ML-1M sample. In the 2D profiles, darker regions correspond to higher dipolar field values. Figures 6(c–f) illustrate the variations in the demagnetization field along lines L1 and L2. A rectangular magnet with a thickness of 25 nm—identical to the one previously considered—was assumed to provide a rough estimate of the FMR shift. The demagnetization factors were computed as $N_x$ = 2.2316, $N_y$ = 0.6135, and $N_z$ = 9.7212. By utilizing the Kittel equation with these numbers, A field variation ($\Delta H$) of approximately 320 Oe is required to produce the expected frequency shift of 1.67 GHz. Notably, the dipolar field variations shown in Figure 6(c–f) for the X- and Y-states range from –60 Oe to –260 Oe, indicating comparable values between the two states. The analytical and simulated dipolar variation parameters show some mismatch because here considered a simple rectangular-shaped nanomagnet instead of a hexagonal one.

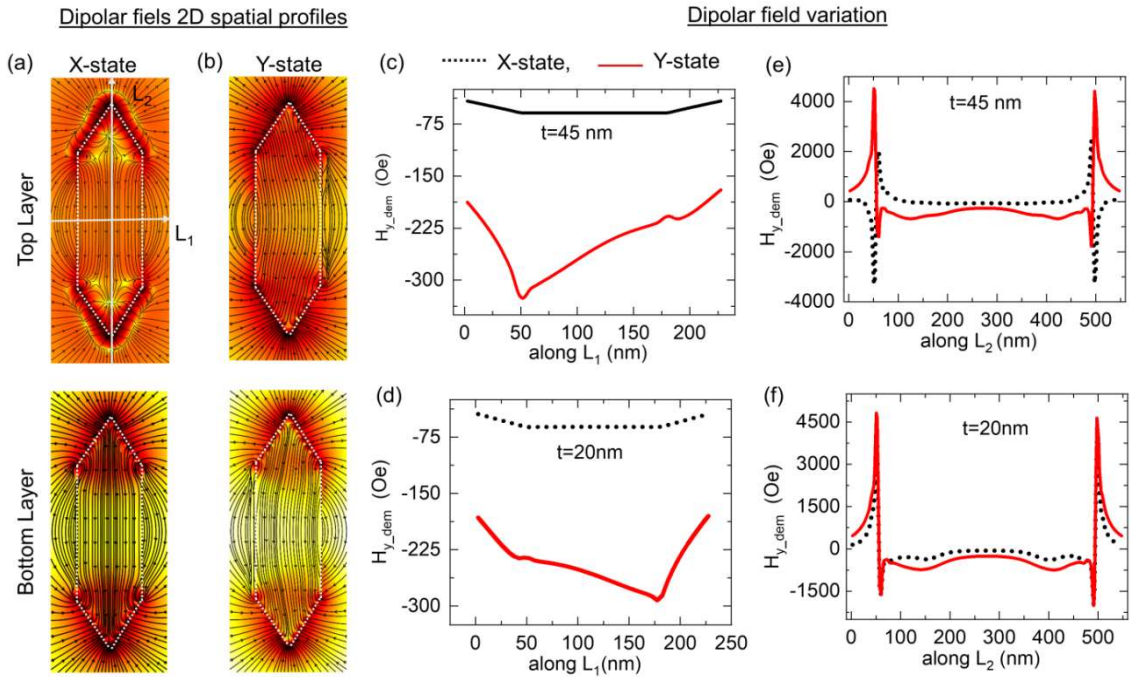

**Figure 5. Dipolar field variation in a multilayer nanomagnet.** (a&b) Two-dimensional maps showing dipolar field lines across the structure, (c&d) spatial distribution of dipolar fields in the top and bottom layers, respectively, and (e&f) line profiles of dipolar field variation along the y-axis for the top and bottom layers.

**C. Multilayer hexagon nanomagnet network systems (ML-HNM-N)**

**Remanent states and microwave properties in dipolar-coupled multilayer nanomagnets**

A designed network systems using multilayer hexagonal nanomagnets with the same dimensions described in the previous section. Two multilayer nanomagnets (ML-2M) were arranged with an interlayer separation of just 30 nm to enable dipolar coupling, forming what I refer to as Network-1 (N1). Similarly, Network-2 (N2) and Network-3 (N3) consist of three (ML-3M) and four (ML-4M) multilayer nanomagnets, respectively. Figure 7(a) presents schematic diagrams of these dipolar-coupled networks. Magnetization reversal was studied using micromagnetic simulations by applying an external magnetic field along the short axis (solid line) and the long axis (dotted line). Interestingly, a two-step magnetic hysteresis loop was observed in all network systems when the field was applied along the long axis, indicating magnetic switching behavior in the multilayer nanomagnets. In contrast, when the magnetic field was applied along the short axis, magnetization reversed in a single step, requiring a higher saturation field of approximately 2200 Oe. The remanent states along the short and long axes are indicated by the star (★) and square (■) symbols, respectively, in the hysteresis loops. Figures 7 (g–h) displays the remanent states of the dipolar-coupled multilayer network systems.

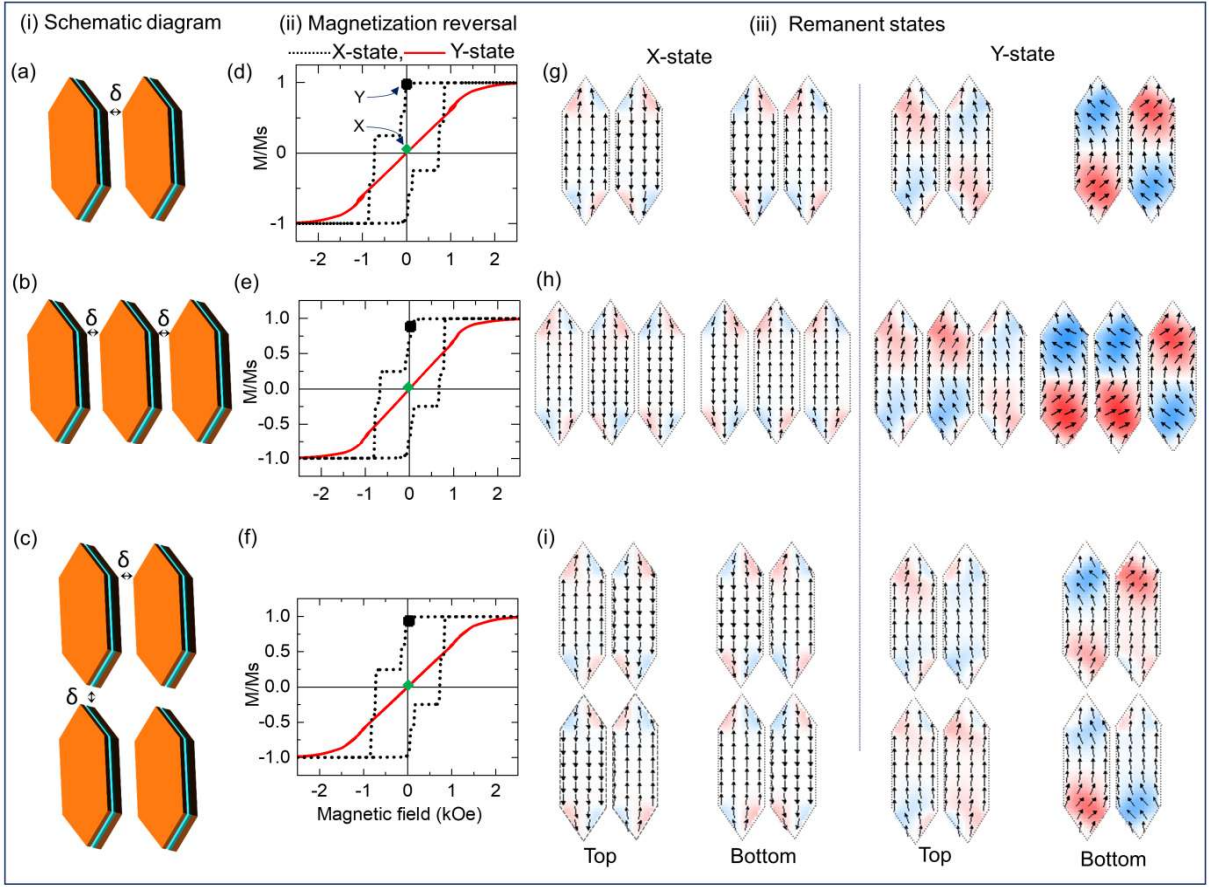

**Figure 6. Dipolar-coupled multilayer nanomagnetic network systems.** (a–c) Schematic representations of the structures, (d–f) simulated magnetization behavior under applied fields, and (g–i) magnetic configurations in the remanent state.

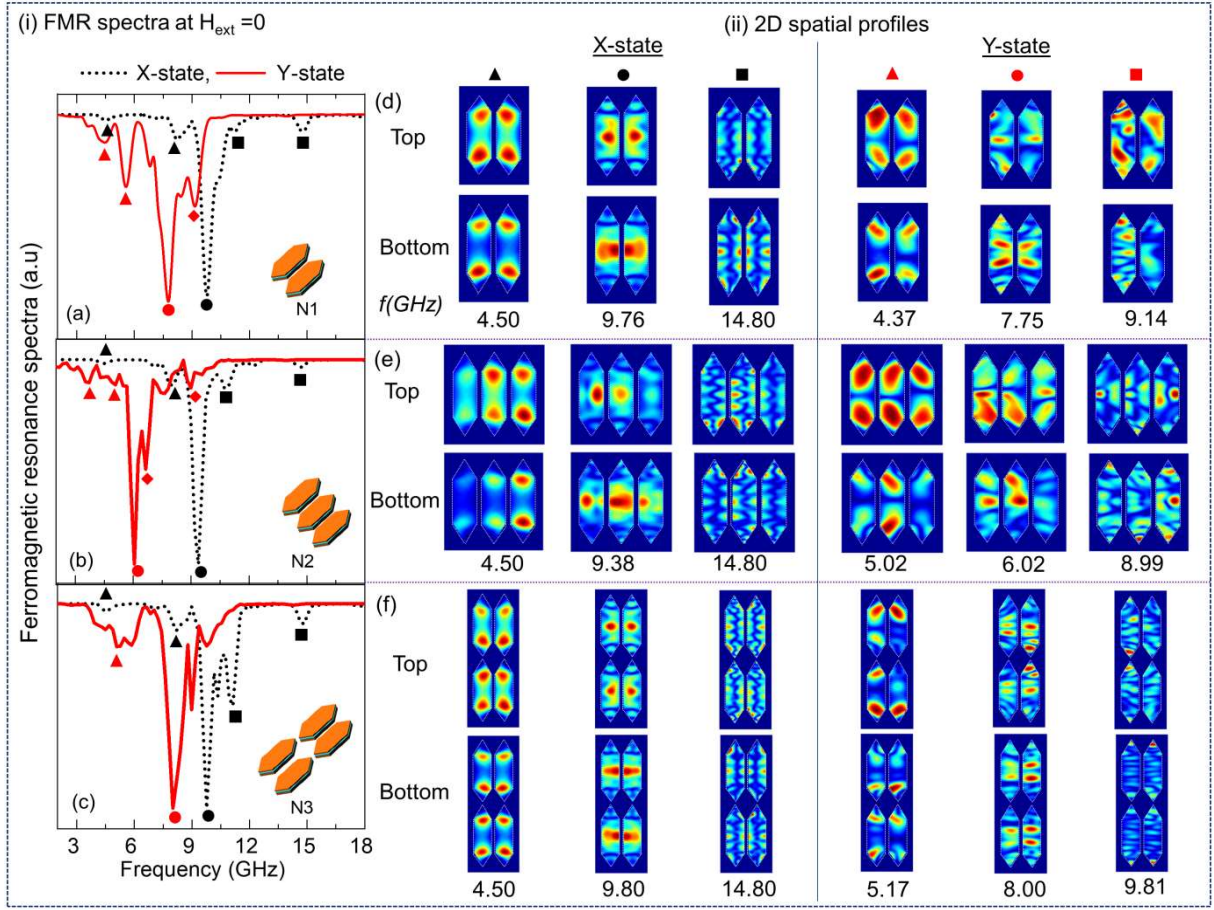

Figure **7. Ferromagnetic resonance of a dipolar-coupled multilayer nanomagnetic networks.** (a) Resonant frequencies of the system, and (b) 2D maps showing the locations of the resonant magnetic activity.

The microwave properties for the remanent states of the network systems are shown in Figure 8 (a–c). The observed frequency range is between 3–16 GHz, which also covering both the X and C bands of the microwave spectrum. The frequency modes for all network systems are listed in the table 1 and compared with those of other systems. Figure 8 (e-f) shows 2D special profile of frequency modes. The center mode ('●') is used to calculate the frequency shift between the X and Y states. A large frequency shift has been observed in the network systems compared to the single-layer and isolated multilayer samples. The observed FMR shift is mainly caused by variations in the dipolar field along the horizontal direction, specifically between different magnetic layers. This significant frequency shift is attributed to strong dipolar interactions and the narrow interlayer spacing in the multilayer structure. A substantial shift of 3.36 GHz was recorded in Network System-2. Dipolar-coupled networks of multilayer nanomagnets with an interlayer separation of 30 nm were examined. The extended structure of ML-1M facilitates additional interlayer interactions, leading to a pronounced variation in the dipolar field, as illustrated in Supplementary Figure S2

*Table 1. Frequency shift of all the nanostructures.*

| Sample | SL-1M | SL-2M | SL-3M | SL-4M | ML-1M | ML-2M | ML-3M | ML-4M |
|---|---|---|---|---|---|---|---|---|
| $\triangle f$ | 800 (MHz) | 300 (MHz) | 618 (MHz) | 200 (MHz) | 1.7 (GHz) | 2.01 (GHz) | 3.36 (GHz) | 1.8 (GHz) |

**D. Ultra-fast switching**

Ultrafast operational capability is governed by the switching speed between the X- and Y-states. This concept was demonstrated using an isolated multilayer nanomagnet, where transitions between X- and Y-states were triggered by applying a pulsed magnetic field along the y-axis or x-axis, respectively. After applying a brief pulse lasting one nanosecond in each case, Figure 9 depicts the time evolution of magnetization. Sub-nanosecond switching capability is indicated by the fact that magnetic switching is well-stabilized within the duration of the pulse field. A one-nanosecond pulse field is indicated by the shaded areas in Figure 9. The pulse field amplitude can be adjusted by altering the sample's geometry and size. For switching from the X-state to the Y-state, an amplitude of 500 Oe is required (Figure 9(a)), whereas the reverse transition from Y-state to X-state requires 950 Oe (Figure 9(b)). The current samples can be used in test and measurement systems or microwave communication equipment, as such devices are commonly used in filters. The inverse spin Hall effect can be used to electronically read out the FMR signal. These samples can be integrated atop cross-point array structures, which allow for directional control of current pulses and the resulting magnetic fields. This configuration enables electrical switching or reconfiguration of the input magnetic states. For dependable performance in integrated systems, a well-defined FMR spectrum with high tunability is essential. In this regard, multilayer structures demonstrate significant FMR shifts, and the number of resonance modes can be tailored by modifying the magnetic layer thicknesses. Unlike patterned arrays, multilayer configurations also support greater device integration density.

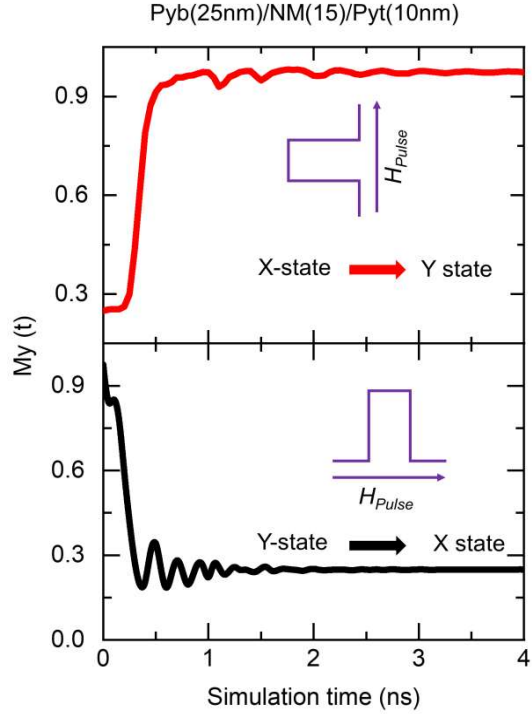

**Figure 8. Ultrafast magnetization switching between stable states in the multilayer nanomagnetic system**. (a) Switching from the X state to the Y state, demonstrating rapid reorientation of magnetic moments under pulse field excitation and (b) reverse switching from the Y state back to the X state, highlighting the system's reconfigurability and dynamic stability.

## IV. SUMMARY

Reprogrammable microwave properties have been demonstrated in a newly designed hexagonal nanomagnet capable of operating without the need for an external bias magnetic field. This study explores three configurations of the nanomagnet: a single-layer structure, an isolated multilayer, and a dipolar-coupled multilayer network. Using micromagnetic simulations, I performed a comprehensive analysis of magnetization reversal, dynamic magnetic behavior, and switching speed. A straightforward field initialization method stabilized two different remanent magnetic states, each characterized by a distinct ferromagnetic resonance (FMR) spectrum. The multilayer configuration exhibited the largest FMR frequency shift (~3.6 GHz), primarily due to variations in dipolar field distributions between the remanent states—reconfigurable on a nanosecond timescale. These results highlight the potential of the hexagonal nanomagnet for use in reprogrammable, low-power, nanoscale microwave devices.

## ACKNOWLEDGMENTS

I gratefully acknowledges the support of the National Post Doctoral Fellowship (PDF/2022/003345) from SERB-DST, Government of India. I sincerely thank Dr. Aarbinda Haldar and Dr. Koteswara Rao Bommisetti for their suggestions and encouragement regarding my career.

SUPPLEMENTARY MATERIALS

Additional information is provided in the supplementary material, as referenced above.

# Supplementary Information

# Micromagnetic Design of Bias-Free Reconfigurable Microwave Properties in $Ni_{80}Fe_{20}$ Hexagonal Nanomagnets

*Krishna Begari*

*Department of Physics, Indian Institute of Technology Tirupati, Tirupati, India*

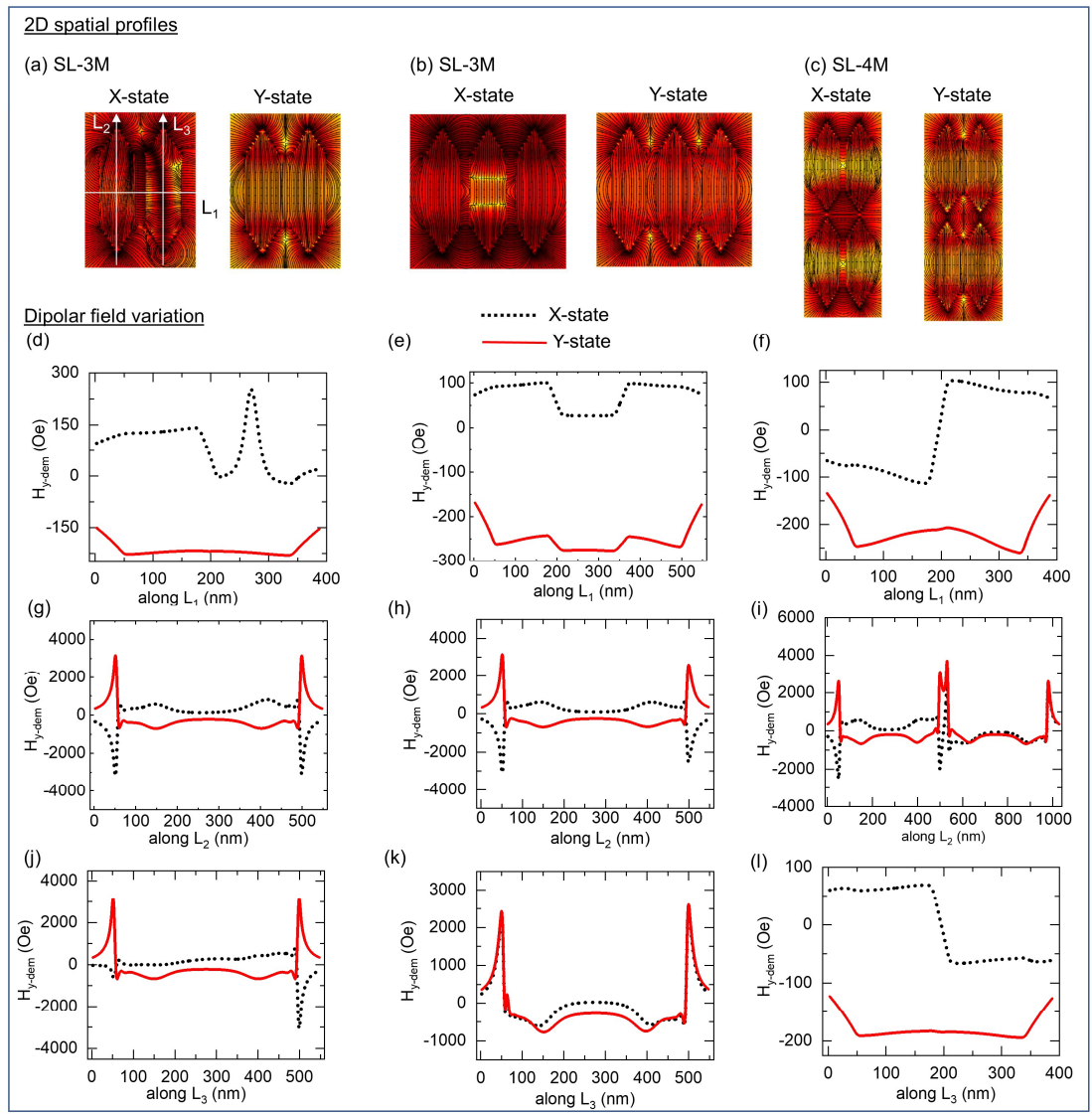

Figure S1. Dipolar field variation in the remanent states of single-layer dipolar-coupled nanomagnets. (a–c) Two-dimensional maps illustrating dipolar field lines in the remanent state. (d–f) Line-scan profiles showing dipolar field variation along the x-axis across the nanomagnets. (g–i) Line scans along the y-axis passing through the center of the first nanomagnet. (j–l) Line scans along the y-axis across the second nanomagnet, capturing its dipolar field distribution.

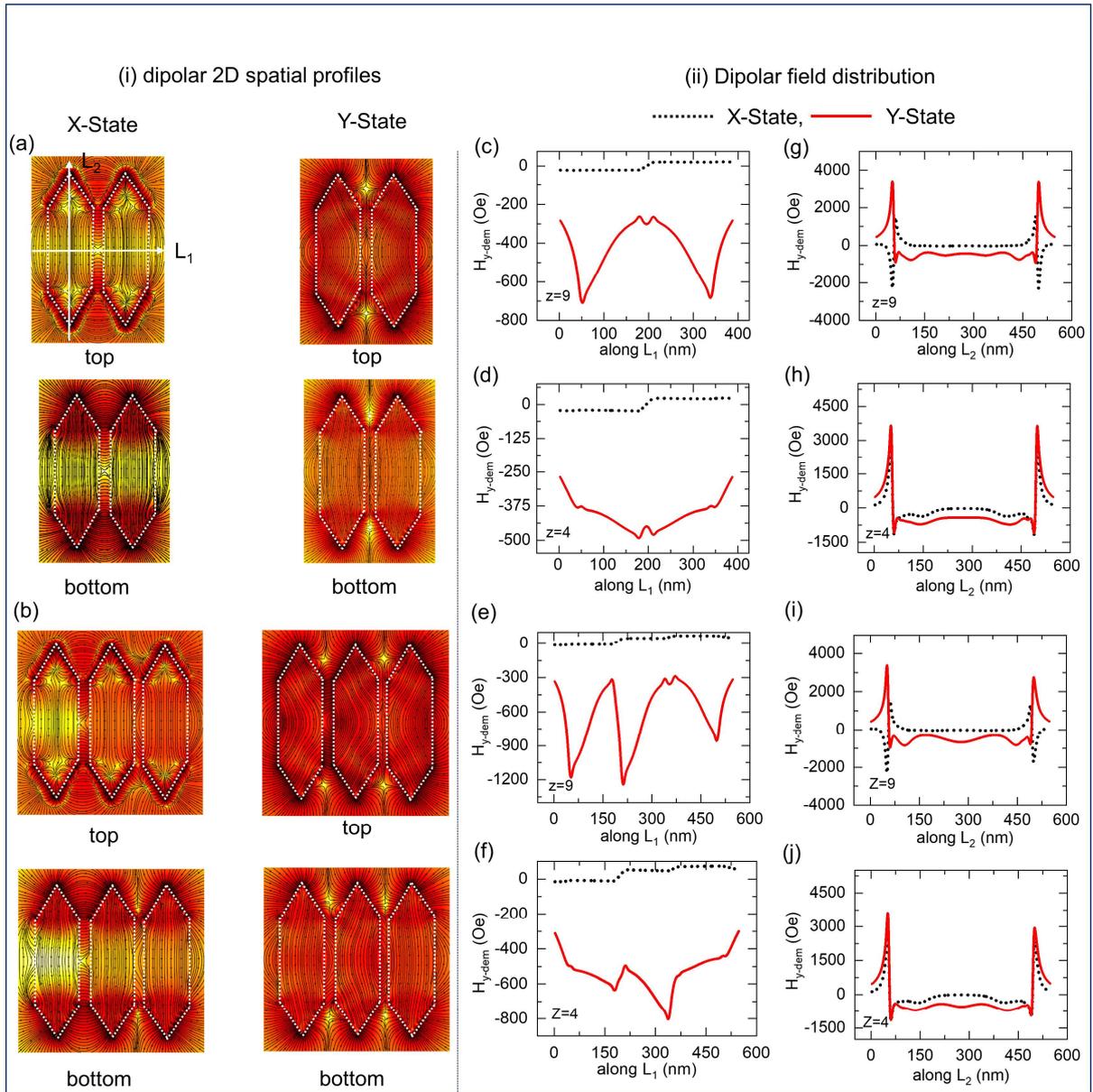

Figure S2 Dipolar field variation in multilayer nanomagnetic network systems. (a, b) Two-dimensional field distribution maps, with dipolar field lines overlaid to illustrate spatial variation. (c–f) Line profiles showing dipolar field variation along the x-axis through the centers of the top and bottom nanomagnets for both X and Y remanent states. (g–j) Corresponding line profiles along the y-axis, highlighting dipolar field differences between the X and Y states.